\documentclass[%
 aip,
 amsmath,amssymb,
 reprint,%
]{revtex4-1}

\usepackage{graphicx}
\usepackage{dcolumn}
\usepackage{bm}

\usepackage[utf8]{inputenc}
\usepackage[T1]{fontenc}
\usepackage{mathptmx}
\usepackage{etoolbox}
\usepackage{xcolor}
\usepackage{siunitx}

\usepackage{upgreek}
\usepackage{verbatim}

\newcommand{\flag}[1]{\textcolor{black}{#1}}
\newcommand{\revcomment}[1]{\textcolor{black}{#1}}
\newcommand{\newtext}[1]{\textcolor{black}{#1}}

\makeatletter
\def\@email#1#2{%
 \endgroup
 \patchcmd{\titleblock@produce}
  {\frontmatter@RRAPformat}
  {\frontmatter@RRAPformat{\produce@RRAP{*#1\href{mailto:#2}{#2}}}\frontmatter@RRAPformat}
  {}{}
}%
\makeatother
\begin{document}

\preprint{AIP/123-QED}
\title[]{Performance of Antenna-based and Rydberg Quantum RF Sensors in the Electrically Small Regime}

\author{K. M. Backes}
\affiliation{Homeland Security Systems Engineering \& Development Institute, an FFRDC operated by The MITRE Corporation for the Department of Homeland Security}
\affiliation{The MITRE Corporation, 7525 Colshire Dr., McLean, VA 22012, USA}
\author{P. K. Elgee}
\affiliation{DEVCOM Army Research Laboratory, 2800 Powder Mill Rd, Adelphi MD 20783, USA}%
\author{K.-J. LeBlanc}
\affiliation{DEVCOM Army Research Laboratory, 2800 Powder Mill Rd, Adelphi MD 20783, USA}%
\author{C. T. Fancher}
\affiliation{Homeland Security Systems Engineering \& Development Institute, an FFRDC operated by The MITRE Corporation for the Department of Homeland Security}
\affiliation{The MITRE Corporation, 7525 Colshire Dr.,  McLean, VA 22012, USA}%
\author{D. H. Meyer}
\affiliation{DEVCOM Army Research Laboratory, 2800 Powder Mill Rd, Adelphi MD 20783, USA}%
\author{P. D. Kunz}
\affiliation{DEVCOM Army Research Laboratory, 2800 Powder Mill Rd, Adelphi MD 20783, USA}%
\author{N. Malvania}
\affiliation{Homeland Security Systems Engineering \& Development Institute, an FFRDC operated by The MITRE Corporation for the Department of Homeland Security}
\affiliation{The MITRE Corporation, 7525 Colshire Dr.,  McLean, VA 22012, USA}%
\author{K. M. Nicolich}
\affiliation{Homeland Security Systems Engineering \& Development Institute, an FFRDC operated by The MITRE Corporation for the Department of Homeland Security}
\affiliation{The MITRE Corporation, 7525 Colshire Dr.,  McLean, VA 22012, USA}%
\author{J. C. Hill}
\affiliation{DEVCOM Army Research Laboratory, 2800 Powder Mill Rd, 
Adelphi MD 20783, USA}%
\author{B. L. Schmittberger Marlow}
\affiliation{Homeland Security Systems Engineering \& Development Institute, an FFRDC operated by The MITRE Corporation for the Department of Homeland Security}
\affiliation{The MITRE Corporation, 7525 Colshire Dr.,  McLean, VA 22012, USA}%
\author{K. C. Cox}\thanks{Corresponding Author, email: kevin.c.cox29.civ@army.mil}
\affiliation{DEVCOM Army Research Laboratory, 2800 Powder Mill Rd, Adelphi MD 20783, USA}%

\begin{abstract}
Rydberg atom electric field sensors are tunable quantum sensors that can perform sensitive radio frequency (RF) measurements. Their qualities have piqued interest at \flag{longer wavelengths} where their small size compares favorably to impedance-matched antennas.  Here, we compare the signal detection sensitivity of cm-scale Rydberg sensors to similarly sized room-temperature electrically small antennas with active and passive receiver backends.  We present and analyze effective circuit models for each sensor type, facilitating a fair sensitivity comparison for cm-scale sensors. We calculate that contemporary Rydberg sensor implementations are less sensitive than unmatched antennas with active amplification. However, we find that idealized Rydberg sensors operating with a maximized atom number and at the standard quantum limit may perform well beyond the capabilities of antenna-based sensors at room temperature, the sensitivities of both lying below typical atmospheric background noise. 
\end{abstract}
\maketitle

The RF spectrum spans more than nine orders of magnitude in frequency, from near-DC to $10^{12}$ Hertz. Due to an increasingly crowded spectrum, modern technologies demand more agility and sensitivity over this large, but finite, usable phase space. Recently, RF sensors based on highly excited atomic Rydberg states have achieved sensitivity across the RF spectrum, demonstrating their broad tunability  \cite{sedlacek_microwave_2012, meyer_assessment_2020-1, jing_atomic_2020, meyer_waveguide-coupled_2021-1, simons_continuous_2021}.  
Rydberg quantum sensors operate based on the quantum physics of atom-photon interactions and \flag{are limited by quantum projection noise instead of internal thermal noise.\cite{wineland_squeezed_1994, mohapatraGiantElectroopticEffect2008, cox_quantum-limited_2018-2}.}
Therefore, they do not obey the same limits as traditional antenna-based sensors.  In this work, we derive and compare idealized and practical sensitivity limits of Rydberg sensors and antenna-based RF sensors with classical electronic readout.

Here, we focus on signal detection in the electrically small regime, where the system size is much smaller than the carrier wavelength $\lambda$.  In this limit, matched antenna-based sensors that efficiently absorb the field energy are often impractical or even impossible to build.  This is because a lossless and matched antenna must be resonant with a large quality factor Q resulting in low bandwidth $\omega_{BW} \sim 1/Q$ \,\cite{chuPhysicalLimitationsOmni1948, harringtonEffectAntennaSize1960}.  Significant research has been conducted on alternative passive and active antenna-based systems to enable more efficient sensing in the electrically small regime. Examples include active non-Foster circuits \cite{sussman-fort_non-foster_2009, linvill_transistor_1953}, optimal matching schemes \cite{best_low_2005, sarabandi_bandwidth_2023}, and active receiver systems \cite{gurses_ultra-sensitive_2021}. The relationship between Rydberg sensors and antenna-based RF sensors, including active circuits, is an area of ongoing study \cite{meyer_assessment_2020-1, fancher_rydberg_2021, santamaria-botello_comparison_2022}.

\newtext{In this work, we first derive the RF detection sensitivities of electrically small antennas coupled to optimal active and passive electronic receiver backends.  Next, we calculate the detection sensitivity of an ideal Rydberg sensor operating at the standard quantum limit (acronymized RSQL) at frequencies below 30~MHz.  We also derive a prediction for the sensitivity achieved by warm Rydberg vapors and electromagnetically induced transparency (REIT sensor) at the same frequencies.  We show that unmatched, active antenna-based circuits outperform REIT sensors and that both can approach atmospheric noise levels. RSQL sensors offer significant opportunity for improvement beyond state-of-the-art sensitivity, which could enable applications such as precision RF metrology in shielded environments.  We present equivalent circuit models that facilitate comparisons between quantum and antenna-based sensors in the electrically small regime, and we conclude by plotting an example comparison for 1~cm-sized sensors in the frequency range of $3\times10^4$ to $3\times10^7$ Hz.  We observe a large performance gap between the sensitivity of current warm-atom based Rydberg sensors and the fundamental sensitivity predicted by the standard quantum limit, motivating further research in Rydberg sensors.}

\begin{figure}
\begin{center}
\includegraphics[scale=1.0]{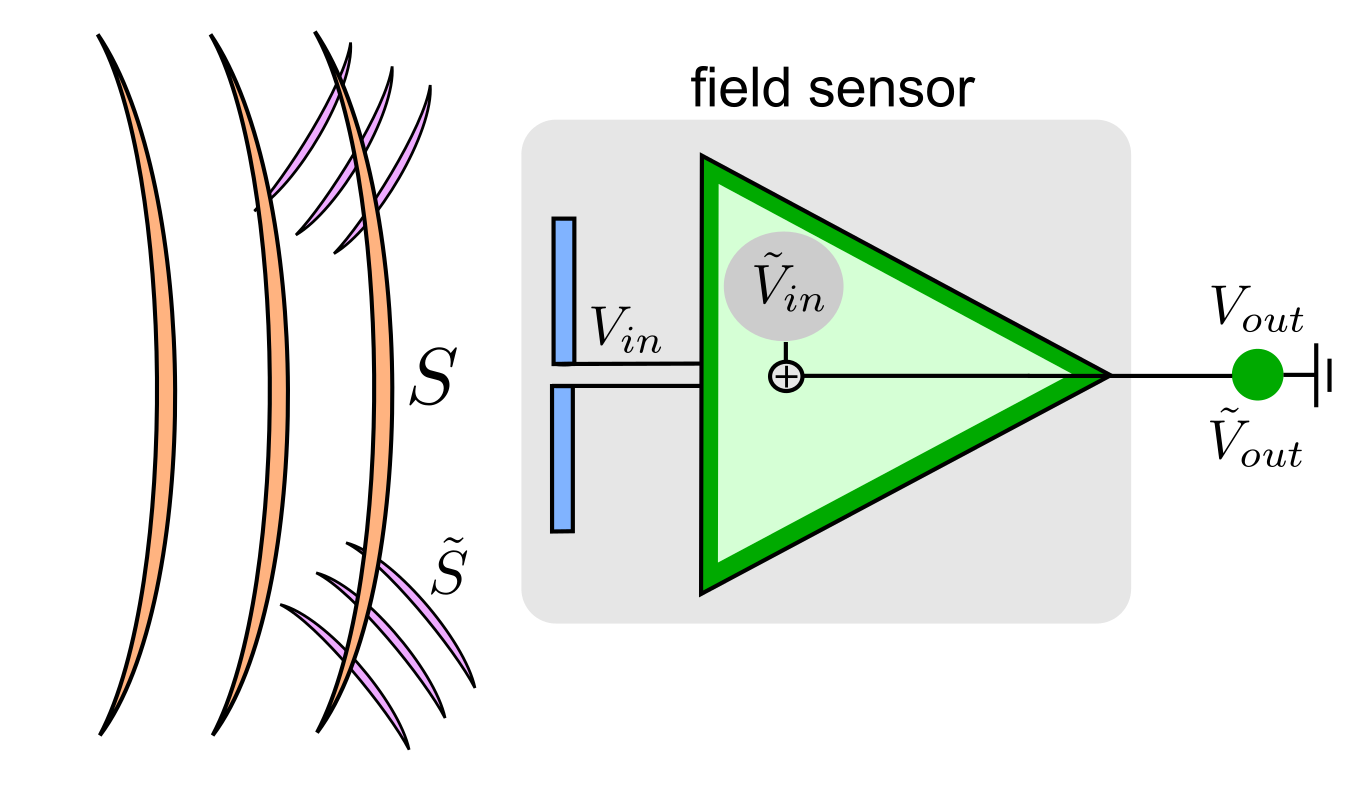}
\caption{Radio-frequency (RF) field detector.  A \flag{plane wave} RF field with Poynting vector amplitude $S$ is absorbed by an antenna or atom represented as a light blue dipole antenna.  \flag{The sensor is also subjected to omnidirectional noise with amplitude $\tilde{S}$}.  The absorbed field is read out by a back end circuit (green).  The minimum detectable RF field is limited by external noise $\tilde{S}$ and equivalent (input-referred) voltage noise inside the sensor $\tilde{V}_{in}$.  }
\label{fig:scenario}
\end{center}
\end{figure}

We begin by calculating the minimum detectable incoming RF-field for each sensor type. We consider an incoming plane-wave RF field as well as external noise characterized by the amplitude of their respective Poynting vectors $S$ and $\tilde{S}$, as depicted in Fig.\,\ref{fig:scenario}.   We use $\sim$ over symbols to represent noise processes which are written in spectral density units.  For example, $\tilde{S}$ has units of W/(m$^2$Hz).

An RF sensor may be limited by either incoming external noise or by internal noise, as shown in Fig.\,\ref{fig:scenario}.  This results in a total signal-to-noise (SNR) density of
\begin{align}
\widetilde{\text{SNR}} \equiv \frac{V_\text{out}^2}{\tilde{V}_\text{out}^2} = \frac{V_\text{in}^2}{\tilde{V}_\text{in}^2 +\tilde{V}_\text{ex}^2} = \frac{\widetilde{\text{SNR}}_\text{in} \, \widetilde{\text{SNR}} _\text{ex}}{\widetilde{\text{SNR}}_\text{in} +\widetilde{\text{SNR}}_\text{ex}},
\end{align}
where $\widetilde{\text{SNR}}_\text{in}$ and $\widetilde{\text{SNR}}_\text{ex}$ are the internal (in) and external (ex) signal-to-noise ratios respectively. At the frequencies considered in this paper and in the presence of atmospheric noise, $\widetilde{\text{SNR}}_\text{ex}$ is often dominant.  But the relative contribution of $\widetilde{\text{SNR}}_\text{ex}$ can be decreased by, for example, constructing a sensor array. $V_{in}$ and $\tilde{V}_{ex}$ are the \flag{Th\'evenin equivalent} voltage amplitudes, and {$V_{in}$ is equivalent to the open circuit voltage for the antenna-based sensors.} 
All signal voltages represent the \newtext{peak} amplitude of oscillating signals at a frequency $\omega$.  The minimum detectable Poynting vector amplitude $S$, with \newtext{spectral density} $\tilde{S}_{min}$, is defined such that $\widetilde{\text{SNR}} = 1$ \flag{in a 1~Hz bandwidth}.

\flag{We treat the scenario depicted in Fig. \ref{fig:scenario}}.  Because a sensor does not differentiate between incoming noise and signal, the only way to improve $\widetilde{\text{SNR}}_\text{ex}$ is to increase the antenna (or Rydberg sensor) directivity.  Electrically small antennas with high directivity are theoretically possible \cite{bouwkamp_problem_1945} although typically impractical.  The antenna gain profile of a Rydberg sensor and a dipole antenna are comparable since we consider atomic transitions where the atom-field interaction is dipolar. Therefore, all sensors considered in this work --- including the Rydberg sensor --- are assumed to have a directivity equal to that of a dipole, namely 3/2.  Thus, all sensors we consider are equivalent with respect to $\widetilde{\text{SNR}}_\text{ex}$.

Given these assumptions, we  narrow focus to the internal signal-to-noise ratio, $\widetilde{\text{SNR}}_{in}$, that quantifies our ability to precisely measure $V_{in}$.  We calculate the input-referred noise voltage $\tilde{V}_\text{in}$.   
$\tilde{V}_{in}$ is also referred to as noise equivalent voltage (NEV) and is related to the noise equivalent input field (NEF) \newtext{$\tilde{E}_{min}$ or $\tilde{B}_{min}$ (for a loop) by applying the appropriate field conversion factor.  }

\begin{figure}
\begin{center}
\includegraphics[scale=0.95]{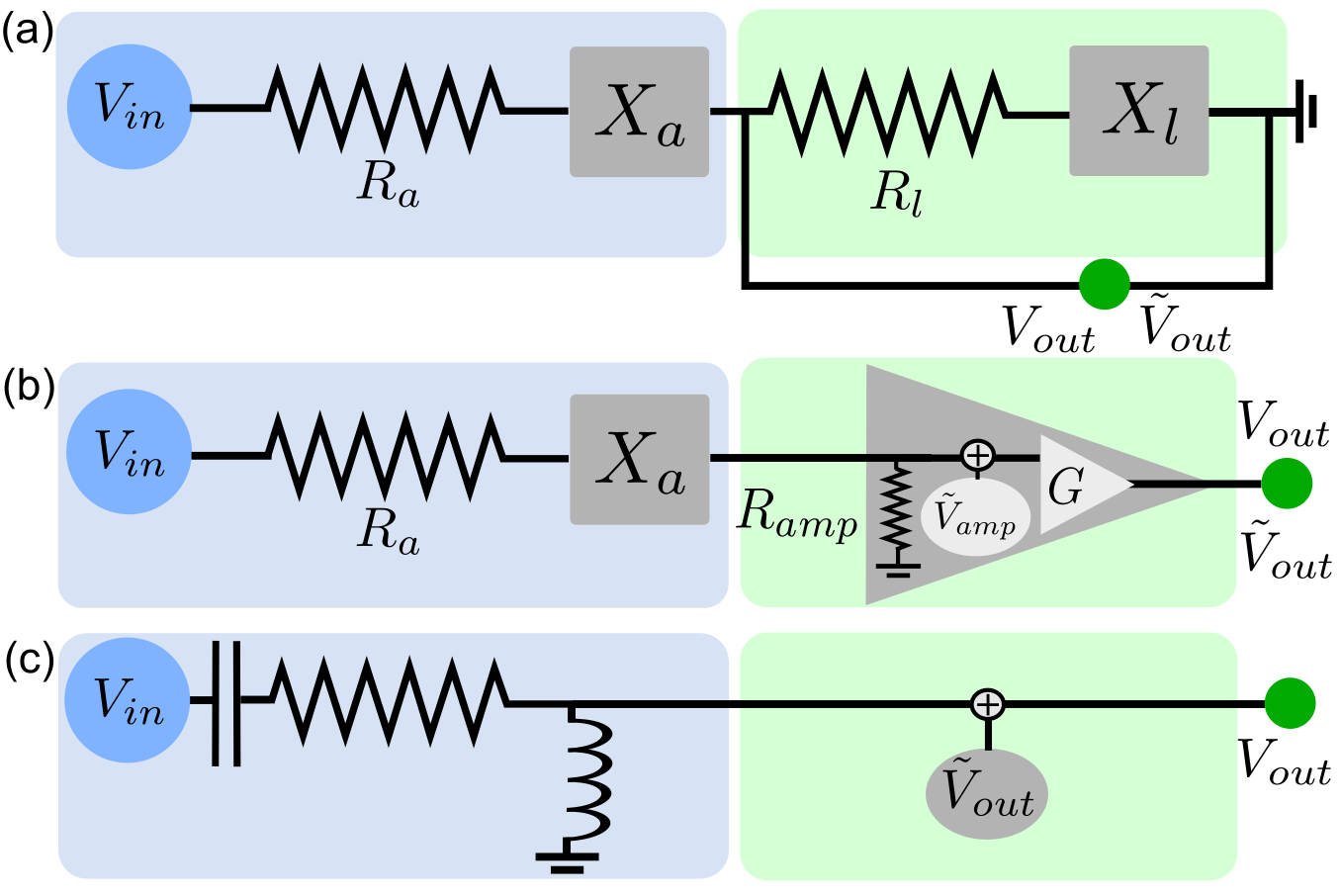}
\caption{Equivalent circuit diagrams for antenna-based and Rydberg RF sensors.  (a) Circuit diagram for a passive small antenna (blue box) coupled to an ideal passive receiver backend (green box).  (b) Equivalent circuit diagram for an actively read out small antenna.  (c) Equivalent circuit for a quantum-limited Rydberg sensor (RSQL sensor).  
}
\label{diagrams}
\end{center}
\end{figure}

First, we represent the antenna (either dipole or loop) as a Th\'evenin equivalent voltage source, a reactance $X_a$, and a resistance $R_a$. \flag{While the antenna resistance $R_a$ is composed of both the physical loss resistance and the radiation resistance, the radiation resistance is negligible at the frequencies considered in this paper.} Figure \ref{diagrams}(a) depicts the equivalent circuit for passive backend with load resistance $R_l$ and load reactance $X_l$.  $V_{out}$ is found by analyzing the RLC circuit with a general result

\begin{align}
V_{out} = V_{in}\gamma(\omega),
\end{align}
where $\gamma(\omega)$ is the magnitude of voltage gain in the RLC circuit\revcomment{\cite{volakisAntennaEngineeringHandbook2007}}, dependent on the carrier frequency $\omega$.
We assume that the noise is dominated by Johnson noise from the resistors $R_a$ and $R_l$ resulting in a total sensor noise of
\begin{align}\label{noise_match}
\tilde{V}^2_\text{out} =  4 k_b T R_{l} \lambda_n^2(\omega)  +4 k_b T R_{a} \gamma_n^2(\omega),
\end{align}
where $T$ is the physical temperature of the antenna and $k_b$ is Boltzmann's constant.  \revcomment{ The two terms correspond to Johnson noise emanating from the load resistance $R_l$ and antenna resistance $R_a$ respectively.  $\lambda_n$ and $\gamma_n$ are the RLC gain experienced by each noise source.  Since antenna noise is present at the circuit input we approximate $\gamma_n \approx \gamma$ and similarly approximate the readout noise gain to one, $\lambda_n \approx 1$.  For accurate analysis of realistic circuits, $\gamma_n$ and $\lambda_n$ can be individually calculated or measured.  Given these approximations, the resulting NEV from Eq. \ref{noise_match} and these considerations is}
\begin{equation}
    \tilde{V}^2_{in} \approx 4 k_b T\bigg(\frac{R_l}{\gamma^2(\omega)} + R_a\bigg).
\end{equation}

For an under-damped resonant circuit that is driven near the resonance frequency $\omega_0$, $\gamma(\omega)$ is approximately Lorentzian with maximum amplitude given by the quality factor $Q$.
In the limit of small load resistance, this yields a fundamental limit to NEV of
\begin{equation}\label{sens_match}
    \tilde{V}^2_{in} \approx 4 k_b T R_a.
\end{equation}
One significant disadvantage of this resonant and matched regime is the low instantaneous bandwidth of the circuit $\omega_\text{BW} \sim \omega_0/Q$.  The maximum of $\omega_\text{BW}$ for a resonant, passive, and lossless system is given by the Chu-Harrington limit \cite{chuPhysicalLimitationsOmni1948, harringtonEffectAntennaSize1960}.  Further, matching a small dipole antenna with large capacitive reactance $X_a$ is often impractical (e.g., requires a large inductor).  

To treat the unmatched case, one may also perform readout using only a resistor, setting the load reactance $X_l$ to zero. 
In this case, \revcomment{the circuit forms a voltage divider with complex gain} $\gamma(\omega) \approx R_l/(R_{l} + X_a)$ \cite{volakisAntennaEngineeringHandbook2007}.  \newtext{For an electrically small dipole antenna, where $X_a$ is large}, the squared NEV is minimized by setting $R_l = X_a$, resulting in
\begin{align}\label{sens_unmatch}
    \tilde{V}^2_{in} \approx 16 k_b T X_a.
\end{align}
This unmatched situation results in a larger NEV than Eq. \ref{sens_match}. 

Next, we consider the ideal active sensor system of Fig.\, \ref{diagrams}(b), characterized by input voltage noise $\tilde{V}_{amp}$, amplifier input impedance $R_{amp}$, and ideal gain $G$. The circuit is limited by the voltage divider formed by $X_a$, $R_a$, and $R_{amp}$.  Noise is contributed by $\tilde{V}_{amp}$ and Johnson noise from $R_a$. The resulting squared NEV is
\begin{align}
\label{sens_active}
\tilde{V}^2_{in} \approx \frac{R^2_{amp}}{(R_{amp}+R_a)^2+X_a^2}( \tilde{V}_{amp}^2 + 4k_b T R_{a}).
\end{align}
If the load impedance $R_{amp}$ is much greater than the antenna reactance $X_a$ and resistance $R_a$, and $\tilde{V}_{amp}$ is small, the active dipole-based sensor attains the ideal sensitivity for the small antenna from Eq. \ref{sens_match}.  But in realistic systems, $\tilde{V}_{amp}$ often dominates.  This scenario also assumes that the active amplifier has enough bandwidth to receive the signal at frequency $\omega$. 

\newtext{Next we tackle the NEV/NEF of a Rydberg sensor.  We first consider the case of an ideal quantum sensor operating at the standard quantum limit (RSQL) \cite{wineland_squeezed_1994, degen_quantum_2017}.  We consider a single Rydberg transition with frequency $\omega_0$ and resonant dipole moment $d$.  We take the incoming field with frequency $\omega$ to be resonant, $\omega = \omega_0$.  } 
We display the Rydberg sensing circuit model in Fig. \ref{diagrams}(c).  
The results of the circuit model yield the same result, by construction, that would be achieved by considering canonical models for quantum sensors \cite{degen_quantum_2017, meyer_assessment_2020-1}.

\newtext{The RSQL sensor is modeled as an RLC resonant circuit in the under-damped regime with resonance frequency $\omega_0$ and quality factor $Q$.  The resonant circuit is driven with an effective (unitless) voltage $V_{in}$, that corresponds with the atom-field interaction.  \revcomment{We define this voltage to be} 
\begin{align}\label{eq:rydberg_vq}
    V_{in} \equiv   \frac{\vec{d} \cdot \vec{E}} {\hbar \omega_0},
\end{align}
where $\vec{d}$ is the dipole moment of the atomic transition. 
\revcomment{$V_{in}$ is defined to map}, for example, to a small angular evolution on the Bloch sphere \cite{degen_quantum_2017}.  The resistor value $R$, and thus the circuit $Q$, is related to the \revcomment{coherent evolution time $\tau$} and decoherence rate $\Gamma$ of the quantum sensor that integrate the effect of the atom-field interaction $V_{in}$,  $Q\equiv \omega_0 \tau e^{-\tau \Gamma}$.  The output of the effective circuit model for the RSQL sensor is $V_{out} = Q V_{in}$.  \flag{ We note that in reality atomic sensors typically operate in the ``rotating wave'' regime, that effectively demodulates the incoming signal against a resonant local oscillator. This phenomenon is not treated explicitly in the circuit model of Fig. \ref{diagrams}(c), since it is not required to achieve the correct sensitivity analysis. }}

The RSQL sensor is limited by noise $\tilde{V}_{out}$ arising from random wave-function collapse of $N$ atoms.  \revcomment{The amplitude of the input voltage $V_{in}$ and noise $\tilde{V}_{out}$ are not physical quantities, but are rather defined to be equal to the known result for quantization noise of a quantum sensor operating at the standard quantum limit \cite{wineland_squeezed_1994, degen_quantum_2017},}

\begin{align}\label{eq:rydberg_snr}
\tilde{V}^2_{out} = \tau/(N).
\end{align}

If the required measurement bandwidth $\omega_{BW}$ is less than twice the atom decoherence rate $\Gamma$, the optimum \revcomment{evolution} time is $\tau = 1/(2 \Gamma)$ \cite{degen_quantum_2017}. Conversely, if $\omega_{BW}$ is greater than $\Gamma$ we choose $\tau = 1/\omega_{BW}$.  This leads to the following NEV with two different regimes,
\begin{align}\label{eq:rydberg_snr}
\tilde{V}^2_{in} =  \begin{cases}
    \frac{\omega_{BW}}{\omega_0^2 N}& (\omega_{BW} \gg \Gamma) \\
     \frac{2 \Gamma}{e \omega_0^2 N}& (\Gamma \gg \omega_{BW}),
 \end{cases}
\end{align}  %
where $e$ is the natural constant.
\newtext{One significant conclusion is that the RSQL sensor is the only sensor considered in this work for which the NEV explicitly depends on sensor bandwidth $\omega_{BW}$.} Assuming the required bandwidth is low ($\omega_{BW} \ll \Gamma$), the NEV may be improved by reducing decoherence $\Gamma$.

\begin{figure}
\begin{center}
\includegraphics[width=\columnwidth]{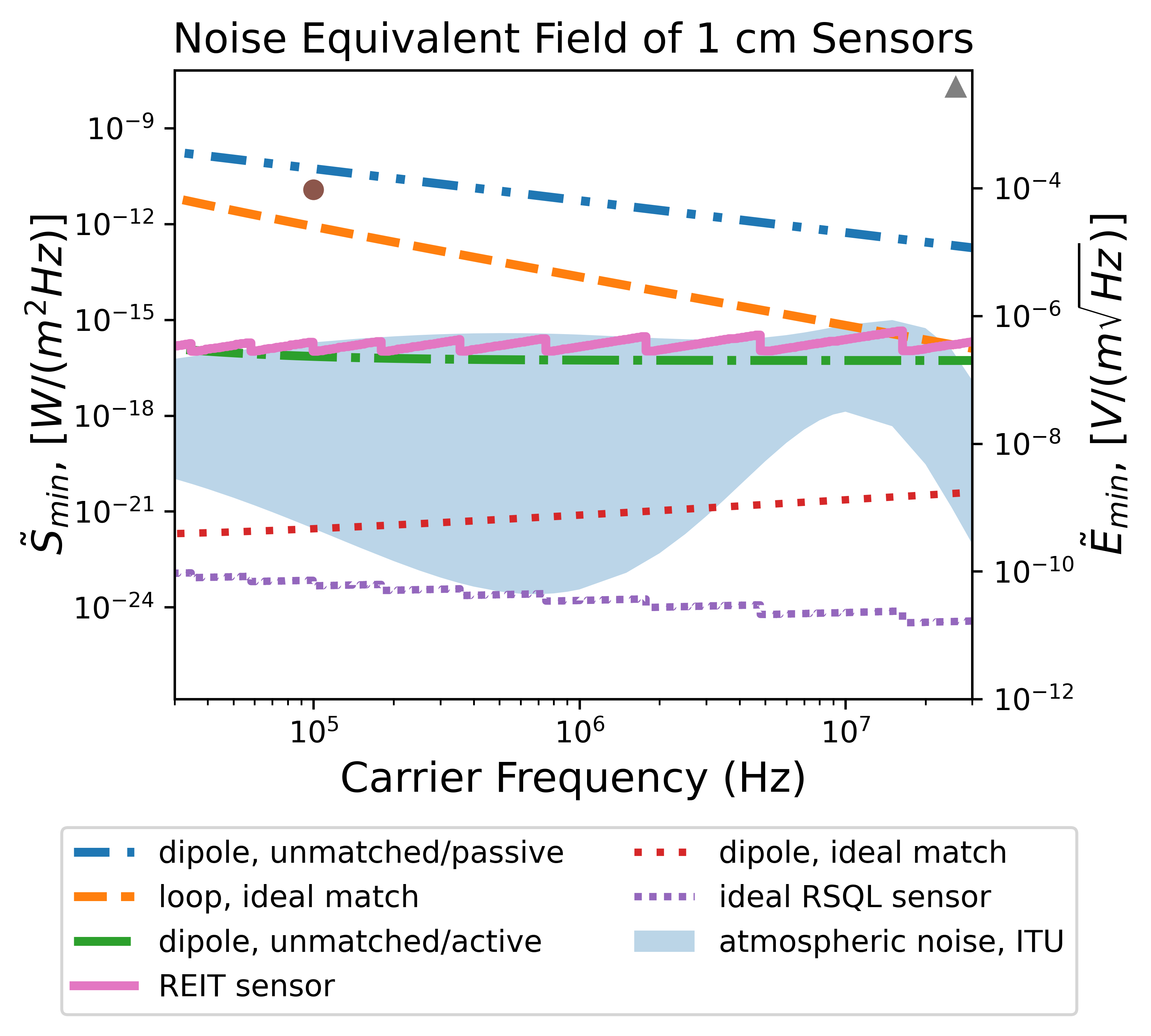}
\caption{Comparison of minimum detectable Poynting vector $\tilde{S}_{min}$ and corresponding minimum detectable E-field $\tilde{E}_{min}$ for Rydberg and antenna-based RF sensors that are 1~cm in scale.  The 10\% - 90\% exceedance band for atmospheric noise referenced to a small dipole antenna is plotted for reference from the ITU Radio Noise recommendations \cite{ITUnoise}.
\newtext{Experimental results from Lei et al.\cite{Lei_100kHz_measurement_2024} and Liu et al. \cite{Liu_30MHz_measurement_2022} are shown as well.}
}
\label{compare}
\end{center}
\end{figure}

We take the treatments described in preceding paragraphs and compare the predicted noise equivalent RF fields in Fig.~\ref{compare}. 
Each of the sensors considered assume a sensing region of approximately 1\,cm.  \flag{Further, to calculate concrete numbers in Fig.~\ref{compare}, several additional assumptions are made including antenna wire dimensions, noise characteristics, and atom number limitations.  The supplementary material is included with figure-generating Python source code to allow these choices to be modified for broad exploration.}

In order to construct the results for antenna-based sensors, we assume that the antenna resistance arises solely from ohmic losses. 
The conductor that makes up the loop and dipole antenna conductor is chosen to be copper with the diameter $a = 1$~mm.  The loop is given a diameter $d =$1~cm, and the dipole is similarly assigned full length $l =$ 1~cm. 
The copper conductor resistivity is \revcomment{$R_s = 1.8\times10^{-8}$\,$\Omega$m} and the magnetic permeability of the loop core is permeability of free space, $\mu$. For the active circuit, we assume the parameters of a low-noise pre-amp with high input impedance, $\tilde{V}_\text{amp} = 1~\text{nV}/ \sqrt\text{Hz}$ and $R_\text{amp} = 100~\text{M}\Omega$. These realistic values are commonly achieved with available amplifiers.

We use Eq.~\ref{sens_match} to add the results for matched/passive antennas to Fig.\,\ref{compare} (orange dash, red dot). Because the single-turn loop that we consider only detects the magnetic field derivative, given by Faraday's law, the loop is significantly less sensitive to the incoming Poynting vector at the frequencies considered here.  However, loop antennas (often with multiple turns and ferrite cores) offer several practical advantages including easily tunable resonance and relatively easy matching.  Conversely, the NEF for the matched dipole starts out frequency independent but rises slightly with frequency due to wire skin depth decreasing with frequency. Although the matched dipole is the most sensitive of the antenna-based sensors, this device suffers significant difficulty in the practicality of lossless matching and the fundamental limitation in bandwidth \revcomment{that results due to the Chu-Harrington limit\cite{chuPhysicalLimitationsOmni1948}}.  

Next, we plot the NEFs for unmatched dipoles with passive (blue dash, Eq.~\ref{sens_unmatch}) and active (green dot-dash) receiver back-ends. The unmatched, passive sensor, similar to what was considered in a previous comparison work \cite{meyer_assessment_2020-1}, is the least sensitive of those considered. The active receiver significantly improves performance by reducing the voltage division effect caused by impedance mismatch due to the amplifier's high input impedance.  For the values chosen,
this active system becomes flat in sensitivity above 100~kHz and is limited solely by the amplifier's input noise.  In the near term, this approach is practical and can surpass the sensitivity of current state-of-the-art Rydberg sensors.

To add RSQL sensors to our comparison, we must apply Eqs.~\ref{eq:rydberg_snr} and ~\ref{eq:rydberg_vq}, which amounts to choosing an appropriate dipole moment, decoherence rate, and atom number. \revcomment{In choosing these values, we assume tightly packed, cold atoms} The result is plotted as purple dots in Fig. \ref{compare}. We apply the following process.  First, we map out the possible Rydberg transitions in rubidium resonant in the VLF-HF bands, restricting ourselves to quantum numbers $l \leq 15$ and $n < 100$\revcomment{, above which operation is non-optimal}.
Using the Atomic Rydberg Calculator (ARC) \cite{sibalic_arc_2017}, we calculate each transition's dipole moment and decoherence rate $\Gamma$ including black body induced decay at room temperature.  The decoherence rate $\Gamma$ is of order $2 \pi \times 1$~kHz over the range of the plot, and exact values may be found (and modified) in the the supplementary material.

Last, we define the atom number by confining ourselves to a $1$~cm$^3$ volume and restricting the density.  The density is chosen to be that which limits van der Waals interactions between atoms to less than a $1$~MHz frequency shift.  This shift is calculated using C$_6$ coefficients from ARC.   The atom number is optimized at each frequency, but is of order $10^8$ over the range of the plot.
This gives us an idealized view of the capability of Rydberg sensors in this frequency regime, yielding a sensitivity $\sim 10^{-11}$~V/(m$\sqrt{\text{Hz}})$ shown as a \revcomment{purple} dotted line in Fig. \ref{compare}.  Numerous effects, including resonant dipole-dipole interactions, plasma formation \cite{weller_interplay_2019}, or other collective effects \cite{young_dissipation-induced_2018} may further limit the ideal sensor.  Much like the line for an ideal matched dipole, the RSQL sensor is likely to be physically unrealizable, but shows a bound of what is possible.  Atomic entanglement may lead to additional improvements \cite{degen_quantum_2017}.  

Finally, we briefly consider REIT sensors and state-of-the-art experimental demonstrations based on warm vapors.  Previous work has calculated the optimal NEF that may be achieved with such sensors \cite{meyer_optimal_2021-1}, and the results are consistent with experimental results \cite{jing_atomic_2020}, resulting in a bounding sensitivity for warm-atom EIT schemes of approximately $10^{-6}~\text{V/(m}\sqrt{\text{Hz}})$.  Recent experiments have further shown that the optimum resonant sensitivity, achieved at microwave frequencies, may be scaled to lower frequencies using high angular momentum states \cite{faconSensitiveElectrometerBased2016, brown_very-high-_2023, berweger_rydberg-state_2023, elgee_satellite_2023}.

\newtext{The estimate of optimal NEF for REIT-based sensors is shown as a pink line in Fig. \ref{compare}.  To calculate this line, we linearly scale the optimum REIT sensitivity  of approximately $10^{-6}~V/(\text{m}\sqrt{\text{Hz}})$ \cite{jing_atomic_2020, meyer_optimal_2021-1} by the dipole transition strengths of available resonant transitions in rubidium 85.   The maximum principal quantum number $n$ is capped at 100 and the angular momentum is capped at quantum number $l = 15$. This scaling process results in the jagged nature of the line.  Nonetheless, the prediction remains of order $10^{-6}$~V/(m$\sqrt{\text{Hz}}$) across the plot range.  We consider this prediction to be optimistic and bounding for REIT-based readout, and practical limitations such as cell shielding \cite{jau_vapor-cell-based_2020} may further limit sensitivity.  We also include two data points corresponding to published experimental results within the plotted range \cite{Liu_30MHz_measurement_2022, Lei_100kHz_measurement_2024}.}

It is valuable to view Fig.\,\ref{compare} as being made up of two regions. The lower two dotted lines in Fig. \ref{compare}, corresponding to the ideal matched dipole and the RSQL sensor are idealizations and both fall beneath typical atmospheric noise levels.  \revcomment{Although these lines will not likely be realized by a physical platform, they nonetheless set an important fundamental bound for the technology.}  In contrast, the lines for the REIT sensor and amplified dipole antenna can likely be achieved in practice, and meet the 10\% exceedance level for atmospheric noise (blue band) that nominally sets a limit for outdoor communications.
Our analysis makes a strong case that unmatched dipole antennas with amplifier readout yield superior performance to what can currently be achieved with REIT sensors. Nonetheless, the large amount of available sensitivity between the two regions of the plot is promising. 

While sensors can approach atmospheric noise levels, sufficient for many applications, Rydberg RF sensors may become useful for reasons apart from their sensitivity.
Recent demonstrations including simultaneous phase and amplitude detection \cite{simons_rydberg_2019}, multi-band reception \cite{meyer_simultaneous_2023}, large dynamic range \cite{anderson_optical_2016}, accurate calibration \cite{anderson_self-calibrated_2021}, Tera-Hertz imaging \cite{downes_full-field_2020}, and sub-thermal detection at microwave frequencies \cite{borowka_continuous_2024} serve as justification and motivation for further research. New sensing schemes that use a larger effective atom number and reach the standard quantum limit (or beyond, by using entanglement) are an important area for future study.  The fact that current Rydberg atom systems are far from the idealized interaction-limited sensitivity is a significant motivation for future research.

\section*{Supplementary Material}

The supplementary material consists of a zipped folder containing a Jupyter notebook and Python source code to re-create Fig. 3 and modify the parameters to create custom comparisons.

\section*{Acknowledgements}
The Authors acknowledge useful discussions with Fredrik Fatemi.

Approved for Public Release; Distribution Unlimited. Public Release Case Number 24-1467. The Homeland Security Act of 2002 (Section 305 of PL 107-296, as codified in 6 U.S.C. 185), herein
referred to as the “Act,” authorizes the Secretary of the Department of Homeland Security (DHS),
acting through the Under Secretary for Science and Technology, to establish one or more federally
funded research and development centers (FFRDCs) to provide independent analysis of
homeland security issues. MITRE Corp. operates the Homeland Security Systems Engineering and
Development Institute (HSSEDI) as an FFRDC for DHS under contract 70RSAT20D00000001.
The HSSEDI FFRDC provides the government with the necessary systems engineering and development
expertise to conduct complex acquisition planning and development; concept exploration,
experimentation and evaluation; information technology, communications and cyber security
processes, standards, methodologies and protocols; systems architecture and integration; quality
and performance review, best practices and performance measures and metrics; and, independent
test and evaluation activities. The HSSEDI FFRDC also works with and supports other federal, state,
local, tribal, public and private sector organizations that make up the homeland security enterprise.
The HSSEDI FFRDC’s research is undertaken by mutual consent with DHS and is organized as
a set of discrete tasks. This report presents the results of research and analysis conducted under:
70RSAT22FR0000021, “DHS Science and Technology Directorate TCD Quantum Information Science Capabilities.”
The results presented in this report do not necessarily reflect official DHS opinion or policy.

The views, opinions and/or findings expressed are those of the authors and should not be interpreted as representing the official views or policies of the Department of Defense or the U.S. Government. 

The authors from The MITRE Corporation also acknowledge support from the MITRE Independent Research and Development Program.

\textbf{Conflicts of Interest}
The authors have no conflicts to disclose.

\textbf{Data Rights}
The data that support the findings of this study are available
from the corresponding author upon reasonable request.

\bibliography{main3}

\begin{thebibliography}{38}%
\makeatletter
\providecommand \@ifxundefined [1]{%
 \@ifx{#1\undefined}
}%
\providecommand \@ifnum [1]{%
 \ifnum #1\expandafter \@firstoftwo
 \else \expandafter \@secondoftwo
 \fi
}%
\providecommand \@ifx [1]{%
 \ifx #1\expandafter \@firstoftwo
 \else \expandafter \@secondoftwo
 \fi
}%
\providecommand \natexlab [1]{#1}%
\providecommand \enquote  [1]{``#1''}%
\providecommand \bibnamefont  [1]{#1}%
\providecommand \bibfnamefont [1]{#1}%
\providecommand \citenamefont [1]{#1}%
\providecommand \href@noop [0]{\@secondoftwo}%
\providecommand \href [0]{\begingroup \@sanitize@url \@href}%
\providecommand \@href[1]{\@@startlink{#1}\@@href}%
\providecommand \@@href[1]{\endgroup#1\@@endlink}%
\providecommand \@sanitize@url [0]{\catcode `\\12\catcode `\$12\catcode `\&12\catcode `\#12\catcode `\^12\catcode `\_12\catcode `\%12\relax}%
\providecommand \@@startlink[1]{}%
\providecommand \@@endlink[0]{}%
\providecommand \url  [0]{\begingroup\@sanitize@url \@url }%
\providecommand \@url [1]{\endgroup\@href {#1}{\urlprefix }}%
\providecommand \urlprefix  [0]{URL }%
\providecommand \Eprint [0]{\href }%
\providecommand \doibase [0]{http://dx.doi.org/}%
\providecommand \selectlanguage [0]{\@gobble}%
\providecommand \bibinfo  [0]{\@secondoftwo}%
\providecommand \bibfield  [0]{\@secondoftwo}%
\providecommand \translation [1]{[#1]}%
\providecommand \BibitemOpen [0]{}%
\providecommand \bibitemStop [0]{}%
\providecommand \bibitemNoStop [0]{.\EOS\space}%
\providecommand \EOS [0]{\spacefactor3000\relax}%
\providecommand \BibitemShut  [1]{\csname bibitem#1\endcsname}%
\let\auto@bib@innerbib\@empty
\bibitem [{\citenamefont {Sedlacek}\ \emph {et~al.}(2012)\citenamefont {Sedlacek}, \citenamefont {Schwettmann}, \citenamefont {K{\"u}bler}, \citenamefont {L{\"o}w}, \citenamefont {Pfau},\ and\ \citenamefont {Shaffer}}]{sedlacek_microwave_2012}%
  \BibitemOpen
  \bibfield  {author} {\bibinfo {author} {\bibfnamefont {J.~A.}\ \bibnamefont {Sedlacek}}, \bibinfo {author} {\bibfnamefont {A.}~\bibnamefont {Schwettmann}}, \bibinfo {author} {\bibfnamefont {H.}~\bibnamefont {K{\"u}bler}}, \bibinfo {author} {\bibfnamefont {R.}~\bibnamefont {L{\"o}w}}, \bibinfo {author} {\bibfnamefont {T.}~\bibnamefont {Pfau}}, \ and\ \bibinfo {author} {\bibfnamefont {J.~P.}\ \bibnamefont {Shaffer}},\ }\bibfield  {title} {\enquote {\bibinfo {title} {Microwave electrometry with {{Rydberg}} atoms in a vapour cell using bright atomic resonances},}\ }\href {\doibase 10.1038/nphys2423} {\bibfield  {journal} {\bibinfo  {journal} {Nature Phys}\ }\textbf {\bibinfo {volume} {8}},\ \bibinfo {pages} {819--824} (\bibinfo {year} {2012})}\BibitemShut {NoStop}%
\bibitem [{\citenamefont {Meyer}\ \emph {et~al.}(2020)\citenamefont {Meyer}, \citenamefont {Castillo}, \citenamefont {Cox},\ and\ \citenamefont {Kunz}}]{meyer_assessment_2020-1}%
  \BibitemOpen
  \bibfield  {author} {\bibinfo {author} {\bibfnamefont {D.~H.}\ \bibnamefont {Meyer}}, \bibinfo {author} {\bibfnamefont {Z.~A.}\ \bibnamefont {Castillo}}, \bibinfo {author} {\bibfnamefont {K.~C.}\ \bibnamefont {Cox}}, \ and\ \bibinfo {author} {\bibfnamefont {P.~D.}\ \bibnamefont {Kunz}},\ }\bibfield  {title} {\enquote {\bibinfo {title} {Assessment of {{Rydberg}} atoms for wideband electric field sensing},}\ }\href {\doibase 10.1088/1361-6455/ab6051} {\bibfield  {journal} {\bibinfo  {journal} {J. Phys. B: At. Mol. Opt. Phys.}\ }\textbf {\bibinfo {volume} {53}},\ \bibinfo {pages} {034001} (\bibinfo {year} {2020})}\BibitemShut {NoStop}%
\bibitem [{\citenamefont {Jing}\ \emph {et~al.}(2020)\citenamefont {Jing}, \citenamefont {Hu}, \citenamefont {Ma}, \citenamefont {Zhang}, \citenamefont {Zhang}, \citenamefont {Xiao},\ and\ \citenamefont {Jia}}]{jing_atomic_2020}%
  \BibitemOpen
  \bibfield  {author} {\bibinfo {author} {\bibfnamefont {M.}~\bibnamefont {Jing}}, \bibinfo {author} {\bibfnamefont {Y.}~\bibnamefont {Hu}}, \bibinfo {author} {\bibfnamefont {J.}~\bibnamefont {Ma}}, \bibinfo {author} {\bibfnamefont {H.}~\bibnamefont {Zhang}}, \bibinfo {author} {\bibfnamefont {L.}~\bibnamefont {Zhang}}, \bibinfo {author} {\bibfnamefont {L.}~\bibnamefont {Xiao}}, \ and\ \bibinfo {author} {\bibfnamefont {S.}~\bibnamefont {Jia}},\ }\bibfield  {title} {\enquote {\bibinfo {title} {Atomic superheterodyne receiver based on microwave-dressed {{Rydberg}} spectroscopy},}\ }\href {\doibase 10.1038/s41567-020-0918-5} {\bibfield  {journal} {\bibinfo  {journal} {Nat. Phys.}\ }\textbf {\bibinfo {volume} {16}},\ \bibinfo {pages} {911--915} (\bibinfo {year} {2020})}\BibitemShut {NoStop}%
\bibitem [{\citenamefont {Meyer}, \citenamefont {Kunz},\ and\ \citenamefont {Cox}(2021)}]{meyer_waveguide-coupled_2021-1}%
  \BibitemOpen
  \bibfield  {author} {\bibinfo {author} {\bibfnamefont {D.~H.}\ \bibnamefont {Meyer}}, \bibinfo {author} {\bibfnamefont {P.~D.}\ \bibnamefont {Kunz}}, \ and\ \bibinfo {author} {\bibfnamefont {K.~C.}\ \bibnamefont {Cox}},\ }\bibfield  {title} {\enquote {\bibinfo {title} {Waveguide-coupled {{Rydberg}} spectrum analyzer from 0 to 20 {{GHz}}},}\ }\href {\doibase 10.1103/PhysRevApplied.15.014053} {\bibfield  {journal} {\bibinfo  {journal} {Phys. Rev. Applied}\ }\textbf {\bibinfo {volume} {15}},\ \bibinfo {pages} {014053} (\bibinfo {year} {2021})},\ \Eprint {http://arxiv.org/abs/2009.14383} {2009.14383 [physics, physics:quant-ph]} \BibitemShut {NoStop}%
\bibitem [{\citenamefont {Simons}\ \emph {et~al.}(2021)\citenamefont {Simons}, \citenamefont {{Artusio-Glimpse}}, \citenamefont {Holloway}, \citenamefont {Imhof}, \citenamefont {Jefferts}, \citenamefont {Wyllie}, \citenamefont {Sawyer},\ and\ \citenamefont {Walker}}]{simons_continuous_2021}%
  \BibitemOpen
  \bibfield  {author} {\bibinfo {author} {\bibfnamefont {M.~T.}\ \bibnamefont {Simons}}, \bibinfo {author} {\bibfnamefont {A.~B.}\ \bibnamefont {{Artusio-Glimpse}}}, \bibinfo {author} {\bibfnamefont {C.~L.}\ \bibnamefont {Holloway}}, \bibinfo {author} {\bibfnamefont {E.}~\bibnamefont {Imhof}}, \bibinfo {author} {\bibfnamefont {S.~R.}\ \bibnamefont {Jefferts}}, \bibinfo {author} {\bibfnamefont {R.}~\bibnamefont {Wyllie}}, \bibinfo {author} {\bibfnamefont {B.~C.}\ \bibnamefont {Sawyer}}, \ and\ \bibinfo {author} {\bibfnamefont {T.~G.}\ \bibnamefont {Walker}},\ }\bibfield  {title} {\enquote {\bibinfo {title} {Continuous radio-frequency electric-field detection through adjacent {{Rydberg}} resonance tuning},}\ }\href {\doibase 10.1103/PhysRevA.104.032824} {\bibfield  {journal} {\bibinfo  {journal} {Phys. Rev. A}\ }\textbf {\bibinfo {volume} {104}},\ \bibinfo {pages} {032824} (\bibinfo {year} {2021})},\ \Eprint {http://arxiv.org/abs/2105.10561} {2105.10561} \BibitemShut {NoStop}%
\bibitem [{\citenamefont {Wineland}\ \emph {et~al.}(1994)\citenamefont {Wineland}, \citenamefont {Bollinger}, \citenamefont {Itano},\ and\ \citenamefont {Heinzen}}]{wineland_squeezed_1994}%
  \BibitemOpen
  \bibfield  {author} {\bibinfo {author} {\bibfnamefont {D.~J.}\ \bibnamefont {Wineland}}, \bibinfo {author} {\bibfnamefont {J.~J.}\ \bibnamefont {Bollinger}}, \bibinfo {author} {\bibfnamefont {W.~M.}\ \bibnamefont {Itano}}, \ and\ \bibinfo {author} {\bibfnamefont {D.~J.}\ \bibnamefont {Heinzen}},\ }\bibfield  {title} {\enquote {\bibinfo {title} {Squeezed atomic states and projection noise in spectroscopy},}\ }\href {\doibase 10.1103/PhysRevA.50.67} {\bibfield  {journal} {\bibinfo  {journal} {Phys. Rev. A}\ }\textbf {\bibinfo {volume} {50}},\ \bibinfo {pages} {67--88} (\bibinfo {year} {1994})}\BibitemShut {NoStop}%
\bibitem [{\citenamefont {Mohapatra}\ \emph {et~al.}(2008)\citenamefont {Mohapatra}, \citenamefont {Bason}, \citenamefont {Butscher}, \citenamefont {Weatherill},\ and\ \citenamefont {Adams}}]{mohapatraGiantElectroopticEffect2008}%
  \BibitemOpen
  \bibfield  {author} {\bibinfo {author} {\bibfnamefont {A.~K.}\ \bibnamefont {Mohapatra}}, \bibinfo {author} {\bibfnamefont {M.~G.}\ \bibnamefont {Bason}}, \bibinfo {author} {\bibfnamefont {B.}~\bibnamefont {Butscher}}, \bibinfo {author} {\bibfnamefont {K.~J.}\ \bibnamefont {Weatherill}}, \ and\ \bibinfo {author} {\bibfnamefont {C.~S.}\ \bibnamefont {Adams}},\ }\bibfield  {title} {\enquote {\bibinfo {title} {A giant electro-optic effect using polarizable dark states},}\ }\href {\doibase 10.1038/nphys1091} {\bibfield  {journal} {\bibinfo  {journal} {Nature Physics}\ }\textbf {\bibinfo {volume} {4}},\ \bibinfo {pages} {890--894} (\bibinfo {year} {2008})}\BibitemShut {NoStop}%
\bibitem [{\citenamefont {Cox}\ \emph {et~al.}(2018)\citenamefont {Cox}, \citenamefont {Meyer}, \citenamefont {Fatemi},\ and\ \citenamefont {Kunz}}]{cox_quantum-limited_2018-2}%
  \BibitemOpen
  \bibfield  {author} {\bibinfo {author} {\bibfnamefont {K.~C.}\ \bibnamefont {Cox}}, \bibinfo {author} {\bibfnamefont {D.~H.}\ \bibnamefont {Meyer}}, \bibinfo {author} {\bibfnamefont {F.~K.}\ \bibnamefont {Fatemi}}, \ and\ \bibinfo {author} {\bibfnamefont {P.~D.}\ \bibnamefont {Kunz}},\ }\bibfield  {title} {\enquote {\bibinfo {title} {Quantum-{{Limited Atomic Receiver}} in the {{Electrically Small Regime}}},}\ }\href {\doibase 10.1103/PhysRevLett.121.110502} {\bibfield  {journal} {\bibinfo  {journal} {Phys. Rev. Lett.}\ }\textbf {\bibinfo {volume} {121}},\ \bibinfo {pages} {110502} (\bibinfo {year} {2018})}\BibitemShut {NoStop}%
\bibitem [{\citenamefont {Chu}(1948)}]{chuPhysicalLimitationsOmni1948}%
  \BibitemOpen
  \bibfield  {author} {\bibinfo {author} {\bibfnamefont {L.~J.}\ \bibnamefont {Chu}},\ }\bibfield  {title} {\enquote {\bibinfo {title} {Physical {{Limitations}} of {{Omni}}-{{Directional Antennas}}},}\ }\href {\doibase 10.1063/1.1715038} {\bibfield  {journal} {\bibinfo  {journal} {Journal of Applied Physics}\ }\textbf {\bibinfo {volume} {19}},\ \bibinfo {pages} {1163--1175} (\bibinfo {year} {1948})}\BibitemShut {NoStop}%
\bibitem [{\citenamefont {Harrington}(1960)}]{harringtonEffectAntennaSize1960}%
  \BibitemOpen
  \bibfield  {author} {\bibinfo {author} {\bibfnamefont {R.~F.}\ \bibnamefont {Harrington}},\ }\bibfield  {title} {\enquote {\bibinfo {title} {Effect of antenna size on gain, bandwidth, and efficiency},}\ }\href@noop {} {\bibfield  {journal} {\bibinfo  {journal} {J. Res. Nat. Bur. Stand}\ }\textbf {\bibinfo {volume} {64}},\ \bibinfo {pages} {1--12} (\bibinfo {year} {1960})}\BibitemShut {NoStop}%
\bibitem [{\citenamefont {{Sussman-Fort}}\ and\ \citenamefont {Rudish}(2009)}]{sussman-fort_non-foster_2009}%
  \BibitemOpen
  \bibfield  {author} {\bibinfo {author} {\bibfnamefont {S.~E.}\ \bibnamefont {{Sussman-Fort}}}\ and\ \bibinfo {author} {\bibfnamefont {R.~M.}\ \bibnamefont {Rudish}},\ }\bibfield  {title} {\enquote {\bibinfo {title} {Non-{{Foster Impedance Matching}} of {{Electrically-Small Antennas}}},}\ }\href {\doibase 10.1109/TAP.2009.2024494} {\bibfield  {journal} {\bibinfo  {journal} {IEEE Transactions on Antennas and Propagation}\ }\textbf {\bibinfo {volume} {57}},\ \bibinfo {pages} {2230--2241} (\bibinfo {year} {2009})}\BibitemShut {NoStop}%
\bibitem [{\citenamefont {Linvill}(1953)}]{linvill_transistor_1953}%
  \BibitemOpen
  \bibfield  {author} {\bibinfo {author} {\bibfnamefont {J.}~\bibnamefont {Linvill}},\ }\bibfield  {title} {\enquote {\bibinfo {title} {Transistor {{Negative-Impedance Converters}}},}\ }\href {\doibase 10.1109/JRPROC.1953.274251} {\bibfield  {journal} {\bibinfo  {journal} {Proceedings of the IRE}\ }\textbf {\bibinfo {volume} {41}},\ \bibinfo {pages} {725--729} (\bibinfo {year} {1953})}\BibitemShut {NoStop}%
\bibitem [{\citenamefont {Best}(2005)}]{best_low_2005}%
  \BibitemOpen
  \bibfield  {author} {\bibinfo {author} {\bibfnamefont {S.}~\bibnamefont {Best}},\ }\bibfield  {title} {\enquote {\bibinfo {title} {Low {{Q}} electrically small linear and elliptical polarized spherical dipole antennas},}\ }\href {\doibase 10.1109/TAP.2004.842600} {\bibfield  {journal} {\bibinfo  {journal} {IEEE Trans. Antennas Propagat.}\ }\textbf {\bibinfo {volume} {53}},\ \bibinfo {pages} {1047--1053} (\bibinfo {year} {2005})}\BibitemShut {NoStop}%
\bibitem [{\citenamefont {Sarabandi}\ and\ \citenamefont {Rao}(2023)}]{sarabandi_bandwidth_2023}%
  \BibitemOpen
  \bibfield  {author} {\bibinfo {author} {\bibfnamefont {K.}~\bibnamefont {Sarabandi}}\ and\ \bibinfo {author} {\bibfnamefont {M.}~\bibnamefont {Rao}},\ }\bibfield  {title} {\enquote {\bibinfo {title} {Bandwidth and {{SNR}} of {{Small Receiving Antennas}}: {{To Match}} or {{Not}} to {{Match}}},}\ }\href {\doibase 10.1109/TAP.2022.3215229} {\bibfield  {journal} {\bibinfo  {journal} {IEEE Transactions on Antennas and Propagation}\ }\textbf {\bibinfo {volume} {71}},\ \bibinfo {pages} {99--104} (\bibinfo {year} {2023})}\BibitemShut {NoStop}%
\bibitem [{\citenamefont {Gurses}, \citenamefont {Whitmore},\ and\ \citenamefont {Cohen}(2021)}]{gurses_ultra-sensitive_2021}%
  \BibitemOpen
  \bibfield  {author} {\bibinfo {author} {\bibfnamefont {B.~V.}\ \bibnamefont {Gurses}}, \bibinfo {author} {\bibfnamefont {K.~T.}\ \bibnamefont {Whitmore}}, \ and\ \bibinfo {author} {\bibfnamefont {M.~B.}\ \bibnamefont {Cohen}},\ }\bibfield  {title} {\enquote {\bibinfo {title} {Ultra-sensitive broadband ``{{AWESOME}}'' electric field receiver for nanovolt low-frequency signals},}\ }\href {\doibase 10.1063/5.0031491} {\bibfield  {journal} {\bibinfo  {journal} {Review of Scientific Instruments}\ }\textbf {\bibinfo {volume} {92}},\ \bibinfo {pages} {024704} (\bibinfo {year} {2021})}\BibitemShut {NoStop}%
\bibitem [{\citenamefont {Fancher}\ \emph {et~al.}(2021)\citenamefont {Fancher}, \citenamefont {Scherer}, \citenamefont {John},\ and\ \citenamefont {Marlow}}]{fancher_rydberg_2021}%
  \BibitemOpen
  \bibfield  {author} {\bibinfo {author} {\bibfnamefont {C.~T.}\ \bibnamefont {Fancher}}, \bibinfo {author} {\bibfnamefont {D.~R.}\ \bibnamefont {Scherer}}, \bibinfo {author} {\bibfnamefont {M.~C.~S.}\ \bibnamefont {John}}, \ and\ \bibinfo {author} {\bibfnamefont {B.~L.~S.}\ \bibnamefont {Marlow}},\ }\bibfield  {title} {\enquote {\bibinfo {title} {Rydberg {{Atom Electric Field Sensors}} for {{Communications}} and {{Sensing}}},}\ }\href {\doibase 10.1109/TQE.2021.3065227} {\bibfield  {journal} {\bibinfo  {journal} {IEEE Transactions on Quantum Engineering}\ }\textbf {\bibinfo {volume} {2}},\ \bibinfo {pages} {1--13} (\bibinfo {year} {2021})}\BibitemShut {NoStop}%
\bibitem [{\citenamefont {{Santamaria-Botello}}\ \emph {et~al.}(2022)\citenamefont {{Santamaria-Botello}}, \citenamefont {Verploegh}, \citenamefont {Bottomley},\ and\ \citenamefont {Popovic}}]{santamaria-botello_comparison_2022}%
  \BibitemOpen
  \bibfield  {author} {\bibinfo {author} {\bibfnamefont {G.}~\bibnamefont {{Santamaria-Botello}}}, \bibinfo {author} {\bibfnamefont {S.}~\bibnamefont {Verploegh}}, \bibinfo {author} {\bibfnamefont {E.}~\bibnamefont {Bottomley}}, \ and\ \bibinfo {author} {\bibfnamefont {Z.}~\bibnamefont {Popovic}},\ }\href {\doibase 10.48550/arXiv.2209.00908} {\enquote {\bibinfo {title} {Comparison of {{Noise Temperature}} of {{Rydberg-Atom}} and {{Electronic Microwave Receivers}}},}\ } (\bibinfo {year} {2022}),\ \Eprint {http://arxiv.org/abs/2209.00908} {2209.00908 [quant-ph]} \BibitemShut {NoStop}%
\bibitem [{\citenamefont {Bouwkamp}\ and\ \citenamefont {{de Bruijn}}(1945)}]{bouwkamp_problem_1945}%
  \BibitemOpen
  \bibfield  {author} {\bibinfo {author} {\bibfnamefont {C.~J.}\ \bibnamefont {Bouwkamp}}\ and\ \bibinfo {author} {\bibfnamefont {N.~G.}\ \bibnamefont {{de Bruijn}}},\ }\bibfield  {title} {\enquote {\bibinfo {title} {The problem of optimum antenna current distribution},}\ }\href@noop {} {\bibfield  {journal} {\bibinfo  {journal} {Philips Research Reports}\ }\textbf {\bibinfo {volume} {1}},\ \bibinfo {pages} {135--158} (\bibinfo {year} {1945})}\BibitemShut {NoStop}%
\bibitem [{\citenamefont {Volakis}(2007)}]{volakisAntennaEngineeringHandbook2007}%
  \BibitemOpen
  \bibfield  {author} {\bibinfo {author} {\bibfnamefont {J.}~\bibnamefont {Volakis}},\ }\href@noop {} {\emph {\bibinfo {title} {Antenna {{Engineering Handbook}}, {{Fourth Edition}}}}},\ \bibinfo {edition} {4th}\ ed.\ (\bibinfo  {publisher} {McGraw-Hill Education},\ \bibinfo {address} {New York},\ \bibinfo {year} {2007})\BibitemShut {NoStop}%
\bibitem [{\citenamefont {Degen}, \citenamefont {Reinhard},\ and\ \citenamefont {Cappellaro}(2017)}]{degen_quantum_2017}%
  \BibitemOpen
  \bibfield  {author} {\bibinfo {author} {\bibfnamefont {C.~L.}\ \bibnamefont {Degen}}, \bibinfo {author} {\bibfnamefont {F.}~\bibnamefont {Reinhard}}, \ and\ \bibinfo {author} {\bibfnamefont {P.}~\bibnamefont {Cappellaro}},\ }\bibfield  {title} {\enquote {\bibinfo {title} {Quantum sensing},}\ }\href {\doibase 10.1103/RevModPhys.89.035002} {\bibfield  {journal} {\bibinfo  {journal} {Rev. Mod. Phys.}\ }\textbf {\bibinfo {volume} {89}},\ \bibinfo {pages} {035002} (\bibinfo {year} {2017})}\BibitemShut {NoStop}%
\bibitem [{\citenamefont {{International Telecommunications Union}}(2022)}]{ITUnoise}%
  \BibitemOpen
  \bibfield  {author} {\bibinfo {author} {\bibnamefont {{International Telecommunications Union}}},\ }\href@noop {} {\enquote {\bibinfo {title} {Recommendation itu-r p.372-16 - radio noise},}\ }\bibinfo {howpublished} {\url{https://www.itu.int/dms_pubrec/itu-r/rec/p/R-REC-P.372-16-202208-I!!PDF-E.pdf}} (\bibinfo {year} {2022})\BibitemShut {NoStop}%
\bibitem [{\citenamefont {Lei}\ and\ \citenamefont {Shi}(2024)}]{Lei_100kHz_measurement_2024}%
  \BibitemOpen
  \bibfield  {author} {\bibinfo {author} {\bibfnamefont {M.}~\bibnamefont {Lei}}\ and\ \bibinfo {author} {\bibfnamefont {M.}~\bibnamefont {Shi}},\ }\href@noop {} {\enquote {\bibinfo {title} {High sensitivity measurement of ulf, vlf and lf fields with rydberg-atom sensor},}\ } (\bibinfo {year} {2024}),\ \Eprint {http://arxiv.org/abs/arXiv:2405.04761} {arXiv:2405.04761} \BibitemShut {NoStop}%
\bibitem [{\citenamefont {Liu}\ \emph {et~al.}(2022)\citenamefont {Liu}, \citenamefont {Zhang}, \citenamefont {Liu}, \citenamefont {Zhang}, \citenamefont {Zhu}, \citenamefont {Gao}, \citenamefont {Guo}, \citenamefont {Ding},\ and\ \citenamefont {Shi}}]{Liu_30MHz_measurement_2022}%
  \BibitemOpen
  \bibfield  {author} {\bibinfo {author} {\bibfnamefont {B.}~\bibnamefont {Liu}}, \bibinfo {author} {\bibfnamefont {L.-H.}\ \bibnamefont {Zhang}}, \bibinfo {author} {\bibfnamefont {Z.-K.}\ \bibnamefont {Liu}}, \bibinfo {author} {\bibfnamefont {Z.-Y.}\ \bibnamefont {Zhang}}, \bibinfo {author} {\bibfnamefont {Z.-H.}\ \bibnamefont {Zhu}}, \bibinfo {author} {\bibfnamefont {W.}~\bibnamefont {Gao}}, \bibinfo {author} {\bibfnamefont {G.-C.}\ \bibnamefont {Guo}}, \bibinfo {author} {\bibfnamefont {D.-S.}\ \bibnamefont {Ding}}, \ and\ \bibinfo {author} {\bibfnamefont {B.-S.}\ \bibnamefont {Shi}},\ }\bibfield  {title} {\enquote {\bibinfo {title} {Highly sensitive measurement of a megahertz rf electric field with a rydberg-atom sensor},}\ }\href {\doibase 10.1103/PhysRevApplied.18.014045} {\bibfield  {journal} {\bibinfo  {journal} {Phys. Rev. Appl.}\ }\textbf {\bibinfo {volume} {18}},\ \bibinfo {pages} {014045} (\bibinfo {year} {2022})}\BibitemShut {NoStop}%
\bibitem [{\citenamefont {{\v S}ibali{\'c}}\ \emph {et~al.}(2017)\citenamefont {{\v S}ibali{\'c}}, \citenamefont {Pritchard}, \citenamefont {Adams},\ and\ \citenamefont {Weatherill}}]{sibalic_arc_2017}%
  \BibitemOpen
  \bibfield  {author} {\bibinfo {author} {\bibfnamefont {N.}~\bibnamefont {{\v S}ibali{\'c}}}, \bibinfo {author} {\bibfnamefont {J.~D.}\ \bibnamefont {Pritchard}}, \bibinfo {author} {\bibfnamefont {C.~S.}\ \bibnamefont {Adams}}, \ and\ \bibinfo {author} {\bibfnamefont {K.~J.}\ \bibnamefont {Weatherill}},\ }\bibfield  {title} {\enquote {\bibinfo {title} {{{ARC}}: {{An}} open-source library for calculating properties of alkali {{Rydberg}} atoms},}\ }\href {\doibase 10.1016/j.cpc.2017.06.015} {\bibfield  {journal} {\bibinfo  {journal} {Computer Physics Communications}\ }\textbf {\bibinfo {volume} {220}},\ \bibinfo {pages} {319--331} (\bibinfo {year} {2017})}\BibitemShut {NoStop}%
\bibitem [{\citenamefont {Weller}\ \emph {et~al.}(2019)\citenamefont {Weller}, \citenamefont {Shaffer}, \citenamefont {Pfau}, \citenamefont {L{\"o}w},\ and\ \citenamefont {K{\"u}bler}}]{weller_interplay_2019}%
  \BibitemOpen
  \bibfield  {author} {\bibinfo {author} {\bibfnamefont {D.}~\bibnamefont {Weller}}, \bibinfo {author} {\bibfnamefont {J.~P.}\ \bibnamefont {Shaffer}}, \bibinfo {author} {\bibfnamefont {T.}~\bibnamefont {Pfau}}, \bibinfo {author} {\bibfnamefont {R.}~\bibnamefont {L{\"o}w}}, \ and\ \bibinfo {author} {\bibfnamefont {H.}~\bibnamefont {K{\"u}bler}},\ }\bibfield  {title} {\enquote {\bibinfo {title} {Interplay between thermal {{Rydberg}} gases and plasmas},}\ }\href {\doibase 10.1103/PhysRevA.99.043418} {\bibfield  {journal} {\bibinfo  {journal} {Phys. Rev. A}\ }\textbf {\bibinfo {volume} {99}},\ \bibinfo {pages} {043418} (\bibinfo {year} {2019})}\BibitemShut {NoStop}%
\bibitem [{\citenamefont {Young}\ \emph {et~al.}(2018)\citenamefont {Young}, \citenamefont {Boulier}, \citenamefont {Magnan}, \citenamefont {Goldschmidt}, \citenamefont {Wilson}, \citenamefont {Rolston}, \citenamefont {Porto},\ and\ \citenamefont {Gorshkov}}]{young_dissipation-induced_2018}%
  \BibitemOpen
  \bibfield  {author} {\bibinfo {author} {\bibfnamefont {J.~T.}\ \bibnamefont {Young}}, \bibinfo {author} {\bibfnamefont {T.}~\bibnamefont {Boulier}}, \bibinfo {author} {\bibfnamefont {E.}~\bibnamefont {Magnan}}, \bibinfo {author} {\bibfnamefont {E.~A.}\ \bibnamefont {Goldschmidt}}, \bibinfo {author} {\bibfnamefont {R.~M.}\ \bibnamefont {Wilson}}, \bibinfo {author} {\bibfnamefont {S.~L.}\ \bibnamefont {Rolston}}, \bibinfo {author} {\bibfnamefont {J.~V.}\ \bibnamefont {Porto}}, \ and\ \bibinfo {author} {\bibfnamefont {A.~V.}\ \bibnamefont {Gorshkov}},\ }\bibfield  {title} {\enquote {\bibinfo {title} {Dissipation-induced dipole blockade and antiblockade in driven {{Rydberg}} systems},}\ }\href {\doibase 10.1103/PhysRevA.97.023424} {\bibfield  {journal} {\bibinfo  {journal} {Phys. Rev. A}\ }\textbf {\bibinfo {volume} {97}},\ \bibinfo {pages} {023424} (\bibinfo {year} {2018})}\BibitemShut {NoStop}%
\bibitem [{\citenamefont {Meyer}\ \emph {et~al.}(2021)\citenamefont {Meyer}, \citenamefont {O'Brien}, \citenamefont {Fahey}, \citenamefont {Cox},\ and\ \citenamefont {Kunz}}]{meyer_optimal_2021-1}%
  \BibitemOpen
  \bibfield  {author} {\bibinfo {author} {\bibfnamefont {D.~H.}\ \bibnamefont {Meyer}}, \bibinfo {author} {\bibfnamefont {C.}~\bibnamefont {O'Brien}}, \bibinfo {author} {\bibfnamefont {D.~P.}\ \bibnamefont {Fahey}}, \bibinfo {author} {\bibfnamefont {K.~C.}\ \bibnamefont {Cox}}, \ and\ \bibinfo {author} {\bibfnamefont {P.~D.}\ \bibnamefont {Kunz}},\ }\bibfield  {title} {\enquote {\bibinfo {title} {Optimal atomic quantum sensing using electromagnetically-induced-transparency readout},}\ }\href {\doibase 10.1103/PhysRevA.104.043103} {\bibfield  {journal} {\bibinfo  {journal} {Phys. Rev. A}\ }\textbf {\bibinfo {volume} {104}},\ \bibinfo {pages} {043103} (\bibinfo {year} {2021})}\BibitemShut {NoStop}%
\bibitem [{\citenamefont {Facon}\ \emph {et~al.}(2016)\citenamefont {Facon}, \citenamefont {Dietsche}, \citenamefont {Grosso}, \citenamefont {Haroche}, \citenamefont {Raimond}, \citenamefont {Brune},\ and\ \citenamefont {Gleyzes}}]{faconSensitiveElectrometerBased2016}%
  \BibitemOpen
  \bibfield  {author} {\bibinfo {author} {\bibfnamefont {A.}~\bibnamefont {Facon}}, \bibinfo {author} {\bibfnamefont {E.-K.}\ \bibnamefont {Dietsche}}, \bibinfo {author} {\bibfnamefont {D.}~\bibnamefont {Grosso}}, \bibinfo {author} {\bibfnamefont {S.}~\bibnamefont {Haroche}}, \bibinfo {author} {\bibfnamefont {J.-M.}\ \bibnamefont {Raimond}}, \bibinfo {author} {\bibfnamefont {M.}~\bibnamefont {Brune}}, \ and\ \bibinfo {author} {\bibfnamefont {S.}~\bibnamefont {Gleyzes}},\ }\bibfield  {title} {\enquote {\bibinfo {title} {A sensitive electrometer based on a {{Rydberg}} atom in a {{Schr{\"o}dinger-cat}} state},}\ }\href {\doibase 10.1038/nature18327} {\bibfield  {journal} {\bibinfo  {journal} {Nature}\ }\textbf {\bibinfo {volume} {535}},\ \bibinfo {pages} {262--265} (\bibinfo {year} {2016})}\BibitemShut {NoStop}%
\bibitem [{\citenamefont {Brown}\ \emph {et~al.}(2023)\citenamefont {Brown}, \citenamefont {Kayim}, \citenamefont {Viray}, \citenamefont {Perry}, \citenamefont {Sawyer},\ and\ \citenamefont {Wyllie}}]{brown_very-high-_2023}%
  \BibitemOpen
  \bibfield  {author} {\bibinfo {author} {\bibfnamefont {R.~C.}\ \bibnamefont {Brown}}, \bibinfo {author} {\bibfnamefont {B.}~\bibnamefont {Kayim}}, \bibinfo {author} {\bibfnamefont {M.~A.}\ \bibnamefont {Viray}}, \bibinfo {author} {\bibfnamefont {A.~R.}\ \bibnamefont {Perry}}, \bibinfo {author} {\bibfnamefont {B.~C.}\ \bibnamefont {Sawyer}}, \ and\ \bibinfo {author} {\bibfnamefont {R.}~\bibnamefont {Wyllie}},\ }\bibfield  {title} {\enquote {\bibinfo {title} {Very-high- and ultrahigh-frequency electric-field detection using high angular momentum {{Rydberg}} states},}\ }\href {\doibase 10.1103/PhysRevA.107.052605} {\bibfield  {journal} {\bibinfo  {journal} {Phys. Rev. A}\ }\textbf {\bibinfo {volume} {107}},\ \bibinfo {pages} {052605} (\bibinfo {year} {2023})}\BibitemShut {NoStop}%
\bibitem [{\citenamefont {Berweger}\ \emph {et~al.}(2023)\citenamefont {Berweger}, \citenamefont {Prajapati}, \citenamefont {{Artusio-Glimpse}}, \citenamefont {Rotunno}, \citenamefont {Brown}, \citenamefont {Holloway}, \citenamefont {Simons}, \citenamefont {Imhof}, \citenamefont {Jefferts}, \citenamefont {Kayim}, \citenamefont {Viray}, \citenamefont {Wyllie}, \citenamefont {Sawyer},\ and\ \citenamefont {Walker}}]{berweger_rydberg-state_2023}%
  \BibitemOpen
  \bibfield  {author} {\bibinfo {author} {\bibfnamefont {S.}~\bibnamefont {Berweger}}, \bibinfo {author} {\bibfnamefont {N.}~\bibnamefont {Prajapati}}, \bibinfo {author} {\bibfnamefont {A.~B.}\ \bibnamefont {{Artusio-Glimpse}}}, \bibinfo {author} {\bibfnamefont {A.~P.}\ \bibnamefont {Rotunno}}, \bibinfo {author} {\bibfnamefont {R.}~\bibnamefont {Brown}}, \bibinfo {author} {\bibfnamefont {C.~L.}\ \bibnamefont {Holloway}}, \bibinfo {author} {\bibfnamefont {M.~T.}\ \bibnamefont {Simons}}, \bibinfo {author} {\bibfnamefont {E.}~\bibnamefont {Imhof}}, \bibinfo {author} {\bibfnamefont {S.~R.}\ \bibnamefont {Jefferts}}, \bibinfo {author} {\bibfnamefont {B.~N.}\ \bibnamefont {Kayim}}, \bibinfo {author} {\bibfnamefont {M.~A.}\ \bibnamefont {Viray}}, \bibinfo {author} {\bibfnamefont {R.}~\bibnamefont {Wyllie}}, \bibinfo {author} {\bibfnamefont {B.~C.}\ \bibnamefont {Sawyer}}, \ and\ \bibinfo {author} {\bibfnamefont {T.~G.}\ \bibnamefont {Walker}},\ }\bibfield  {title} {\enquote {\bibinfo {title} {Rydberg-{{State
  Engineering}}: {{Investigations}} of {{Tuning Schemes}} for {{Continuous Frequency Sensing}}},}\ }\href {\doibase 10.1103/PhysRevApplied.19.044049} {\bibfield  {journal} {\bibinfo  {journal} {Phys. Rev. Appl.}\ }\textbf {\bibinfo {volume} {19}},\ \bibinfo {pages} {044049} (\bibinfo {year} {2023})}\BibitemShut {NoStop}%
\bibitem [{\citenamefont {Elgee}\ \emph {et~al.}(2023)\citenamefont {Elgee}, \citenamefont {Hill}, \citenamefont {LeBlanc}, \citenamefont {Ko}, \citenamefont {Kunz}, \citenamefont {Meyer},\ and\ \citenamefont {Cox}}]{elgee_satellite_2023}%
  \BibitemOpen
  \bibfield  {author} {\bibinfo {author} {\bibfnamefont {P.~K.}\ \bibnamefont {Elgee}}, \bibinfo {author} {\bibfnamefont {J.~C.}\ \bibnamefont {Hill}}, \bibinfo {author} {\bibfnamefont {K.-J.~E.}\ \bibnamefont {LeBlanc}}, \bibinfo {author} {\bibfnamefont {G.~D.}\ \bibnamefont {Ko}}, \bibinfo {author} {\bibfnamefont {P.~D.}\ \bibnamefont {Kunz}}, \bibinfo {author} {\bibfnamefont {D.~H.}\ \bibnamefont {Meyer}}, \ and\ \bibinfo {author} {\bibfnamefont {K.~C.}\ \bibnamefont {Cox}},\ }\bibfield  {title} {\enquote {\bibinfo {title} {Satellite radio detection via dual-microwave {{Rydberg}} spectroscopy},}\ }\href {\doibase 10.1063/5.0158150} {\bibfield  {journal} {\bibinfo  {journal} {Applied Physics Letters}\ }\textbf {\bibinfo {volume} {123}},\ \bibinfo {pages} {084001} (\bibinfo {year} {2023})}\BibitemShut {NoStop}%
\bibitem [{\citenamefont {Jau}\ and\ \citenamefont {Carter}(2020)}]{jau_vapor-cell-based_2020}%
  \BibitemOpen
  \bibfield  {author} {\bibinfo {author} {\bibfnamefont {Y.-Y.}\ \bibnamefont {Jau}}\ and\ \bibinfo {author} {\bibfnamefont {T.}~\bibnamefont {Carter}},\ }\bibfield  {title} {\enquote {\bibinfo {title} {Vapor-{{Cell-Based Atomic Electrometry}} for {{Detection Frequencies}} below 1 {{kHz}}},}\ }\href {\doibase 10.1103/PhysRevApplied.13.054034} {\bibfield  {journal} {\bibinfo  {journal} {Phys. Rev. Applied}\ }\textbf {\bibinfo {volume} {13}},\ \bibinfo {pages} {054034} (\bibinfo {year} {2020})},\ \Eprint {http://arxiv.org/abs/2002.04145} {2002.04145} \BibitemShut {NoStop}%
\bibitem [{\citenamefont {Simons}\ \emph {et~al.}(2019)\citenamefont {Simons}, \citenamefont {Haddab}, \citenamefont {Gordon},\ and\ \citenamefont {Holloway}}]{simons_rydberg_2019}%
  \BibitemOpen
  \bibfield  {author} {\bibinfo {author} {\bibfnamefont {M.~T.}\ \bibnamefont {Simons}}, \bibinfo {author} {\bibfnamefont {A.~H.}\ \bibnamefont {Haddab}}, \bibinfo {author} {\bibfnamefont {J.~A.}\ \bibnamefont {Gordon}}, \ and\ \bibinfo {author} {\bibfnamefont {C.~L.}\ \bibnamefont {Holloway}},\ }\bibfield  {title} {\enquote {\bibinfo {title} {A {{Rydberg}} atom-based mixer: {{Measuring}} the phase of a radio frequency wave},}\ }\href {\doibase 10.1063/1.5088821} {\bibfield  {journal} {\bibinfo  {journal} {Appl. Phys. Lett.}\ }\textbf {\bibinfo {volume} {114}},\ \bibinfo {pages} {114101} (\bibinfo {year} {2019})}\BibitemShut {NoStop}%
\bibitem [{\citenamefont {Meyer}\ \emph {et~al.}(2023)\citenamefont {Meyer}, \citenamefont {Hill}, \citenamefont {Kunz},\ and\ \citenamefont {Cox}}]{meyer_simultaneous_2023}%
  \BibitemOpen
  \bibfield  {author} {\bibinfo {author} {\bibfnamefont {D.~H.}\ \bibnamefont {Meyer}}, \bibinfo {author} {\bibfnamefont {J.~C.}\ \bibnamefont {Hill}}, \bibinfo {author} {\bibfnamefont {P.~D.}\ \bibnamefont {Kunz}}, \ and\ \bibinfo {author} {\bibfnamefont {K.~C.}\ \bibnamefont {Cox}},\ }\bibfield  {title} {\enquote {\bibinfo {title} {Simultaneous {{Multiband Demodulation Using}} a {{Rydberg Atomic Sensor}}},}\ }\href {\doibase 10.1103/PhysRevApplied.19.014025} {\bibfield  {journal} {\bibinfo  {journal} {Phys. Rev. Appl.}\ }\textbf {\bibinfo {volume} {19}},\ \bibinfo {pages} {014025} (\bibinfo {year} {2023})}\BibitemShut {NoStop}%
\bibitem [{\citenamefont {Anderson}\ \emph {et~al.}(2016)\citenamefont {Anderson}, \citenamefont {Miller}, \citenamefont {Raithel}, \citenamefont {Gordon}, \citenamefont {Butler},\ and\ \citenamefont {Holloway}}]{anderson_optical_2016}%
  \BibitemOpen
  \bibfield  {author} {\bibinfo {author} {\bibfnamefont {D.~A.}\ \bibnamefont {Anderson}}, \bibinfo {author} {\bibfnamefont {S.~A.}\ \bibnamefont {Miller}}, \bibinfo {author} {\bibfnamefont {G.}~\bibnamefont {Raithel}}, \bibinfo {author} {\bibfnamefont {J.~A.}\ \bibnamefont {Gordon}}, \bibinfo {author} {\bibfnamefont {M.~L.}\ \bibnamefont {Butler}}, \ and\ \bibinfo {author} {\bibfnamefont {C.~L.}\ \bibnamefont {Holloway}},\ }\bibfield  {title} {\enquote {\bibinfo {title} {Optical {{Measurements}} of {{Strong Microwave Fields}} with {{Rydberg Atoms}} in a {{Vapor Cell}}},}\ }\href {\doibase 10.1103/PhysRevApplied.5.034003} {\bibfield  {journal} {\bibinfo  {journal} {Phys. Rev. Appl.}\ }\textbf {\bibinfo {volume} {5}},\ \bibinfo {pages} {034003} (\bibinfo {year} {2016})}\BibitemShut {NoStop}%
\bibitem [{\citenamefont {Anderson}, \citenamefont {Sapiro},\ and\ \citenamefont {Raithel}(2021)}]{anderson_self-calibrated_2021}%
  \BibitemOpen
  \bibfield  {author} {\bibinfo {author} {\bibfnamefont {D.~A.}\ \bibnamefont {Anderson}}, \bibinfo {author} {\bibfnamefont {R.~E.}\ \bibnamefont {Sapiro}}, \ and\ \bibinfo {author} {\bibfnamefont {G.}~\bibnamefont {Raithel}},\ }\bibfield  {title} {\enquote {\bibinfo {title} {A {{Self-Calibrated SI-Traceable Rydberg Atom-Based Radio Frequency Electric Field Probe}} and {{Measurement Instrument}}},}\ }\href {\doibase 10.1109/TAP.2021.3060540} {\bibfield  {journal} {\bibinfo  {journal} {IEEE Transactions on Antennas and Propagation}\ }\textbf {\bibinfo {volume} {69}},\ \bibinfo {pages} {5931--5941} (\bibinfo {year} {2021})}\BibitemShut {NoStop}%
\bibitem [{\citenamefont {Downes}\ \emph {et~al.}(2020)\citenamefont {Downes}, \citenamefont {MacKellar}, \citenamefont {Whiting}, \citenamefont {Bourgenot}, \citenamefont {Adams},\ and\ \citenamefont {Weatherill}}]{downes_full-field_2020}%
  \BibitemOpen
  \bibfield  {author} {\bibinfo {author} {\bibfnamefont {L.~A.}\ \bibnamefont {Downes}}, \bibinfo {author} {\bibfnamefont {A.~R.}\ \bibnamefont {MacKellar}}, \bibinfo {author} {\bibfnamefont {D.~J.}\ \bibnamefont {Whiting}}, \bibinfo {author} {\bibfnamefont {C.}~\bibnamefont {Bourgenot}}, \bibinfo {author} {\bibfnamefont {C.~S.}\ \bibnamefont {Adams}}, \ and\ \bibinfo {author} {\bibfnamefont {K.~J.}\ \bibnamefont {Weatherill}},\ }\bibfield  {title} {\enquote {\bibinfo {title} {Full-{{Field Terahertz Imaging}} at {{Kilohertz Frame Rates Using Atomic Vapor}}},}\ }\href {\doibase 10.1103/PhysRevX.10.011027} {\bibfield  {journal} {\bibinfo  {journal} {Phys. Rev. X}\ }\textbf {\bibinfo {volume} {10}},\ \bibinfo {pages} {011027} (\bibinfo {year} {2020})},\ \Eprint {http://arxiv.org/abs/1903.01308} {1903.01308} \BibitemShut {NoStop}%
\bibitem [{\citenamefont {Bor{\'o}wka}\ \emph {et~al.}(2024)\citenamefont {Bor{\'o}wka}, \citenamefont {Pylypenko}, \citenamefont {Mazelanik},\ and\ \citenamefont {Parniak}}]{borowka_continuous_2024}%
  \BibitemOpen
  \bibfield  {author} {\bibinfo {author} {\bibfnamefont {S.}~\bibnamefont {Bor{\'o}wka}}, \bibinfo {author} {\bibfnamefont {U.}~\bibnamefont {Pylypenko}}, \bibinfo {author} {\bibfnamefont {M.}~\bibnamefont {Mazelanik}}, \ and\ \bibinfo {author} {\bibfnamefont {M.}~\bibnamefont {Parniak}},\ }\bibfield  {title} {\enquote {\bibinfo {title} {Continuous wideband microwave-to-optical converter based on room-temperature {{Rydberg}} atoms},}\ }\href {\doibase 10.1038/s41566-023-01295-w} {\bibfield  {journal} {\bibinfo  {journal} {Nat. Photon.}\ }\textbf {\bibinfo {volume} {18}},\ \bibinfo {pages} {32--38} (\bibinfo {year} {2024})}\BibitemShut {NoStop}%
\end{thebibliography}%


\end{document}